\documentclass[fleqn,usenatbib]{mnras}

\usepackage{newtxtext,newtxmath}

\usepackage[T1]{fontenc}

\DeclareRobustCommand{\VAN}[3]{#2}
\let\VANthebibliography\thebibliography
\def\thebibliography{\DeclareRobustCommand{\VAN}[3]{##3}\VANthebibliography}

\usepackage{graphicx}	
\usepackage{amsmath}	
\usepackage{caption}

\newcommand\msun{M_{\odot}}
\newcommand\rsun{R_{\odot}}

\newcommand\mdot{\dot{M}}
\newcommand\msunyr{M_{\odot}\, {\rm yr^{-1}}}

\title[Outbursts in embedded protostars]{
Observational signatures of
outside-in accretion bursts in embedded protostars}

\author[A. T. Masley et al.]{
August T. Masley,$^{1,2}$
Lee Hartmann$^{1}$
\\
$^{1}$Department of Astronomy, University of Michigan, 1085 South University Avenue, Ann Arbor, MI 48015, USA\\
$^{2}$Ritter Astrophysical Research Center, Department of Physics and Astronomy, University of Toledo, Toledo, OH 43606, USA
}

\date{Accepted XXX. Received YYY; in original form ZZZ}

\pubyear{2025}

\begin{document}
\label{firstpage}
\pagerange{\pageref{firstpage}--\pageref{lastpage}}
\maketitle

\begin{abstract}
Optical and infrared surveys have detected
increasing numbers of disc accretion outbursts
in young stars. Some models of these FU Ori-type 
events predict that the outburst should
start at near- to mid-infrared wavelengths before an optical rise is detected,
and this lag between infrared and optical
bursts has been observed in at least two systems. Detecting and characterizing infrared
precursors can constrain the outburst trigger region, and thus help identify the
mechanism producing the outburst. 
However,
because FU Ori objects are generally young and
usually embedded in dusty protostellar envelopes,
it is not clear whether or how well
such infrared precursors can be detected in the presence 
of strong envelope extinction. 
To explore this question, we combine time-dependent outburst models of the inner disc with an outer dusty disc and protostellar envelope,
and calculate the resulting spectral energy distributions 
(SEDs) using the radiative transfer code RADMC3D. 
We find that, for envelope mass infall rates 
$\gtrsim 10^{-5}\msunyr \, (r_c/{\rm 30\, au})^{-1/2}$, where
$r_c$ is a characteristic
inner radius for the infalling envelope,
the infrared precursor is only apparent in the SED when viewed along an outflow cavity. At other inclinations, the precursor is 
most easily distinguished
with limited envelope extinction at infall rates 
$\lesssim 10^{-6} \msunyr
\, (r_c/{\rm 30 \, au})^{-1/2}$.
We also show that far-infrared and submm/mm
monitoring can enable
the indirect detection of
precursor evolution long before the optical
outburst, emphasizing the potential
of long-wavelength monitoring for studying the earliest stages of
protostar formation.

\end{abstract}




\section{Introduction}
\label{sec:intro}

The study of accretion variability in young
stars is being increasingly seen as providing important clues to disc physics and further insight into the processes by which stars accumulate their masses \cite[e.g,][]{fischer23}. Of special interest are FU
Ori objects \citep{hartmann1996}, many of which exhibit some of the largest accretion outbursts.
A common model for these bursts is thermal
triggering of the magnetorotational instability (MRI) \citep{balbus1991}, which results in a large increase in the local viscosity and/or
angular momentum transport, and a corresponding
increase in accretion
\citep{armitage01,zhu10a,zhu10b,bae13,bae14}.

While some FU Ori outbursts may require initiation in the innermost disc \citep{bell94}, 
models that reproduce rapid, large rises in
optical light invoke triggering at radii
$\sim 0.1 - 1$~au and subsequent propagation
inward. 
This outside-in burst picture predicts the appearance of an infrared precursor before
the optical outburst becomes evident, and
this has been observed in some (relatively low-luminosity) systems
\citep{hillenbrand2018,szegedi2020}.
The detection of such infrared precursors
and measurement of the lag timescales for the
subsequent optical brightening, can provide
important constraints on where the outbursts
are initiated and how they propagate,
which in turn are crucial to identify outburst
mechanisms
\citep{cleaver2023}. 
Moreover, detection of
infrared precursors can help
identify new FU Ori objects
\citep{contreras25}.

However, as FU Ori objects are often embedded in their dusty envelopes, 
a combination of extinction and thermal emission can make it
difficult to identify
an object in outburst.
\cite{fischer24} explored
how observations at 
protostellar envelopes at mid-infrared to sub-mm wavelengths could
be used to identify outbursts in systems whose
central regions are heavily obscured.
Modeling the accretion bursts as increases in luminosity
of the central star, Fischer et al. found that
observations at mid- to
far-infrared wavelengths
provided the best estimates
of changes in protostellar accretion rates. 

Our purpose here is
different,
in that we focus the detectability
of non-steady disc temperature distributions and outburst temporal
evolution on spectral energy distributions (SEDs). 
We adopt one-dimensional, time-dependent models of disc outbursts from the work of
\cite{cleaver2023} as input to radiative transfer 
models of protostellar systems with infalling envelopes to explore the sensitivity of the SEDs to different stages of outside-in accretion
outbursts.
The models assume that a large $\alpha$ viscosity occurs when the disc internal temperature exceeds $\sim 1300$~K, to simulate triggering of the magnetorotational instability (MRI) due to thermal ionization in an otherwise cold and inert disc
\citep{armitage01,zhu10a,zhu10b}.
The envelopes are irradiated by the accreting disc, in contrast to other treatments that often
assume a spherical symmetry for the input radiation
field. The initial signature of
the outburst occurs appears at infrared wavelengths, with the optical outburst coming later. We find that this infrared precursor will
be clearly detectable when observed along outflow cavities, but may
not be apparent when observing through the extincting envelope, depending upon the mass infall rate, detailed envelope geometry, and viewing inclination.
However, even when the infrared
peak is not detectable, monitoring light curves 
at mid- to far-infrared and submm/mm
wavelengths can provide important 
constraints on outburst properties and
mechanisms.

We discuss our methods in \S \ref{sec:meth}, and show our results in \S \ref{sec:results}. 
The implications of our results along with comparisons
with the work of \cite{fischer24} 
are explored
in \S \ref{sec:discussion}, 
and our conclusions are summarized in \S \ref{sec:conc}.

\section{Methods}
\label{sec:meth}

\subsection{Radiative transfer}

We use the software package RADMC3D \citep{dullemond2012} 
to calculate the temperature structure and emergent spectrum of our disc $+$ envelope models.
RADMC3D can accurately return a self-consistent temperature distribution based on the Monte Carlo method of \citet{bjorkman2001}. We make use
of
RADMC3D's ability to include multiple dust species, each with its own density distribution and opacity table. In addition, we use the ``heat source'' option 
to account for the radiation produced by our luminous outbursting disc, as described in 
\S \ref{sec:innerdisc}.

\subsection{Density distribution}

The two main components of our model density distributions are an envelope and a disc. The envelope is described using the density distribution for a
collapsing and rotating singular sphere \citep{TSC1984}, 
with the inner region described by the solutions of
\citet{ulrich1976} and \citet{cassen1981},
\begin{equation}
    \rho_{env} = \frac{\dot{M}}{4 \pi (GMr^3)^{1/2}}\left(1 + \frac{\cos{\theta}}{\cos{\theta_\circ}}\right)^{-1/2}\left(\frac{\cos{\theta}}{\cos{\theta_\circ}} + \frac{2\cos^2{\theta_\circ}}{r/r_c}\right)^{-1}
	\label{eq:envdens}
\end{equation}
Here $\dot{M}$ is the infall rate, $M$ is the central body's mass, $r$ and $\theta$ are the radial distance and polar angle in spherical coordinates, $\theta_\circ$ is the polar angle of a given streamline at infinity, and $r_c$ is the centrifugal radius of the envelope.
By using the mass infall rate as a parameter, we specify the stellar mass to fix the density needed
for the radiative transfer solution.
To simulate an outflow
cavity, we set the density
to zero
for streamlines with
$\theta_{\circ} \leq \theta_{out}$. We take
the asymptotic cavity opening
angle to be $\theta_{out} = 35^{\circ}$ unless otherwise specified.

The inner and outer disc regions
require different treatment
because the temperature distribution at small radii
is driven by the disc accretion,
while the outer cold dusty disc
is passively irradiated by
the inner regions. 
At radii less than $10$~AU,
the disc is assumed to be geometrically thin and optically thick, with the radiative flux as a function of cylindrical radius taken from the 1-D time-dependent outburst models of \cite{cleaver2023}. 
The surface density $\Sigma$ is kept
constant for $R < 10$~au at a value
chosen to be continuous with the
outer disk density distribution.
The reason for this is to limit the optical depth of the inner
disc as discussed in~\ref{sec:innerdisc}.
At larger radii the disc is assumed to have
a radial surface density distribution with a truncated power law in cylindrical radius $R$ used in many works. Here we specifically adopt
\begin{equation}
    \Sigma(R) = \Sigma_0 \left(\frac{R}{R_0}\right)^{-1} \exp\left({-\frac{R}{R_T}}\right) \,.
	\label{eq:surfdens}
\end{equation}
\citep[e.g.,][]{hartmann98}.
Similarly, a typical assumption of vertical
isothermality is used for the structure in the
$z$ direction,
\begin{equation}
    \rho_{disc} = \rho_0 (R) \,\exp \left({\frac{-z^2}{2H^2}}\right)\,,
	\label{eq:diskdens}
\end{equation}
where $\rho_0$ is the midplane density
and $H$ is the scale height.
We assume a radial dependence of the scale height
$H = H_0 (R/R_0)^{5/4}$ which implies a disc
temperature $T \propto R^{-1/2}$. We do not iterate the disc temperature and vertical hydrostatic equilibrium for consistency, as our focus here is not
on the geometrical disc structure.
As in \cite{han2023}, the transition between the disc and envelope density distributions is set where the 
disc thermal pressure $P_{disc} = \rho_{disc}\ c_s^2$, matches the ram pressure of the infalling envelope, $ \rho_{env}\ v_{\theta}^2$, where $c_s$ is the sound speed and $v_\theta$ is the velocity
component approximately perpendicular to the disc surface.

\subsection{Dust opacities}

We assume that two primary dust species, olivine silicates and graphite,
are present in the envelope and outer disc,
with compositions given by \citet{draine1984}. The opacities for these grains are calculated using the Mie scattering methods of \citet{alessio2001}, based on the values of \citet{doschner1995} and \citet{draine1984}.

For the envelope, we use these dust opacities 
assuming the usual power law distribution
of dust as a function of radius $N(a) \propto a^{-3.5}$, with a maximum size $a_{max} = 0.3\, \mu$m typical of interstellar medium diffuse regions. For the disc, we adopt $a_{max} = 1$~mm, 
to allow for the settling of large dust as required in many systems \citep{dalessio01,villenave20}.

\subsection{Inner disc in outbursts}
\label{sec:innerdisc}

The temperature structure of the inner disc
is taken from the time-dependent simulations computed by \cite{cleaver2023}.
These one-dimensional models
assume that the optically-thick
disc is heated by an $\alpha$ viscosity which turns on smoothly between $1200$~K to 1300~K,
to simulate activation of the MRI by thermal ionization.\footnote{Some calculations have suggested lower activation temperatures $\sim 800 - 1000$~K \citep{desch15}, but changing this parameter
would have little effect on our findings.}
The SEDs were calculated by summing up the flux at each radial annulus assuming blackbody radiation 
for regions where dust opacity dominates, and stellar atmosphere
models when dust is sublimated and the dominant opacity is from the gas.

During outbursts the inner disc becomes the main source of radiation rather than
the central star.
We therefore used
the internal heat source method of RADMC3D, which allows us to input values of viscous heating directly. We take the energy fluxes at each
radius from the calculations of \cite{cleaver2023}
and inject that heating into two
grid cells spanning the midplane. This method allows us to generate
a central source of radiation that has a more
appropriate angular distribution of emergent
intensities than that of a spherical source
(i.e., discs radiate preferentially perpendicular
to their surface). It also results in a radiating source with a more realistic spatial distribution than if we
assumed a central radiation source. 
The values of the heat flux at each disc radius were
taken from the snapshots of the time-dependent outburst
models (i.e., from the effective temperature distributions), which can differ drastically from that of a steady
accretion disc.

In the heat source method, RADMC3D assumes that the opacity is that of dust. 
The time-dependent outburst models of \cite{cleaver2023} use gas and dust Rosseland mean opacities internally and stellar model atmospheres to compute the emergent spectra.
As we do not use gaseous opacities for the inner
disc, we must use some other prescription
to compute the disc SED. Furthermore,
because accretion rates
in the outburst models can be very high (peak values of
$10^{-4} \msunyr$), central temperatures in the inner
disc at peak outburst can 
approach $5 \times 10^4$~K, due to both the large viscous heating and the large optical depths that trap
the radiation
(Figure \ref{fig:jacobtempplot}).
We quickly found that adopting typical dust
opacities (ignoring sublimation)
and surface densities from the time-dependent
simulations similarly resulted in very
high internal temperatures that RADMC3D 
had great difficulty in handling.
We therefore limit the surface density within $R < 10$~AU
and apply a so-called "gray opacity" (adopting a constant absorptive opacity of 0.1 ${\rm g\ cm^{-2}}$)
while simultaneously ensuring sufficient optical depth (in excess of $\tau \sim 10$), 
to ensure the appropriate emission at the
disc surface while limiting the central temperatures
and the number of scatterings required for the energy to escape. 

Adopting a gray, temperature-independent opacity 
results in
each disc annulus at 
$R < 10$~au emitting as a blackbody, depending upon the local effective temperature. 
We therefore do not reproduce the
detailed spectral features of the
SEDs in \cite{cleaver2023}, but this
is unimportant for the purposes of
irradiating the envelope and outer disc,
as our method reproduces the overall shape
of the disc SEDs reasonably well.

\section{Results}
\label{sec:results}

\subsection{Model parameters}\label{sec:params}

\begin{table}
    \caption{Model parameters}
    \begin{tabular}{ |c|c|c|c|c|c| }
        \hline
        Model & Outburst  & Infall rate & $r_c$&
        Cavity & Outburst age\\
         & type &  ($\msunyr$) &(au) &$\theta_{out}$  &(yr) \\
        \hline
        1 & none & $10^{-5}$ & 30 & $35^{\circ}$ &  none\\
        2A & FU & $10^{-5}$ & 30 &" &  50\\
        2B & "  & $10^{-5}$ & 30 &" & 70\\
         2B' & "  & $10^{-5}$ & 30 &
         $ 10^{\circ}$ & 70\\
        2C & "  & $10^{-5}$ & 30 &$35^{\circ}$ & 85\\
        2D & " & $10^{-5}$ & 30 & " & 100 \\
        3 & "  & $10^{-6}$ & 30 & " & 70\\
        4A & " & $10^{-5}$ & 3 & " & 70 \\
        4B & " & $10^{-5}$ & 30 & "& 70 \\
        5A & Gaia & $10^{-5}$ & 30 & " &  0 \\
        5B & "  & $10^{-5}$ & 30 & " & 3 \\
        5C & " & $10^{-5}$ & 30 & " & 5 \\
        5D & " & $10^{-6}$ & 30 & " & 
        0 \\
        5E & " & $10^{-6}$ & 30 & " &  
        3 \\
        5F & " & $10^{-6}$ & 30 & " &  
        5 \\
\hline
    \end{tabular}

Outburst type 
is either
the strong, FU Ori-like outburst (FU) or the 
a weaker, Gaia 17bpi-like outburst (Gaia) cases
from \cite{cleaver2023}.

    \label{tab:modelparams}
 \end{table}

\begin{figure}\centering
    \includegraphics[width=\columnwidth]{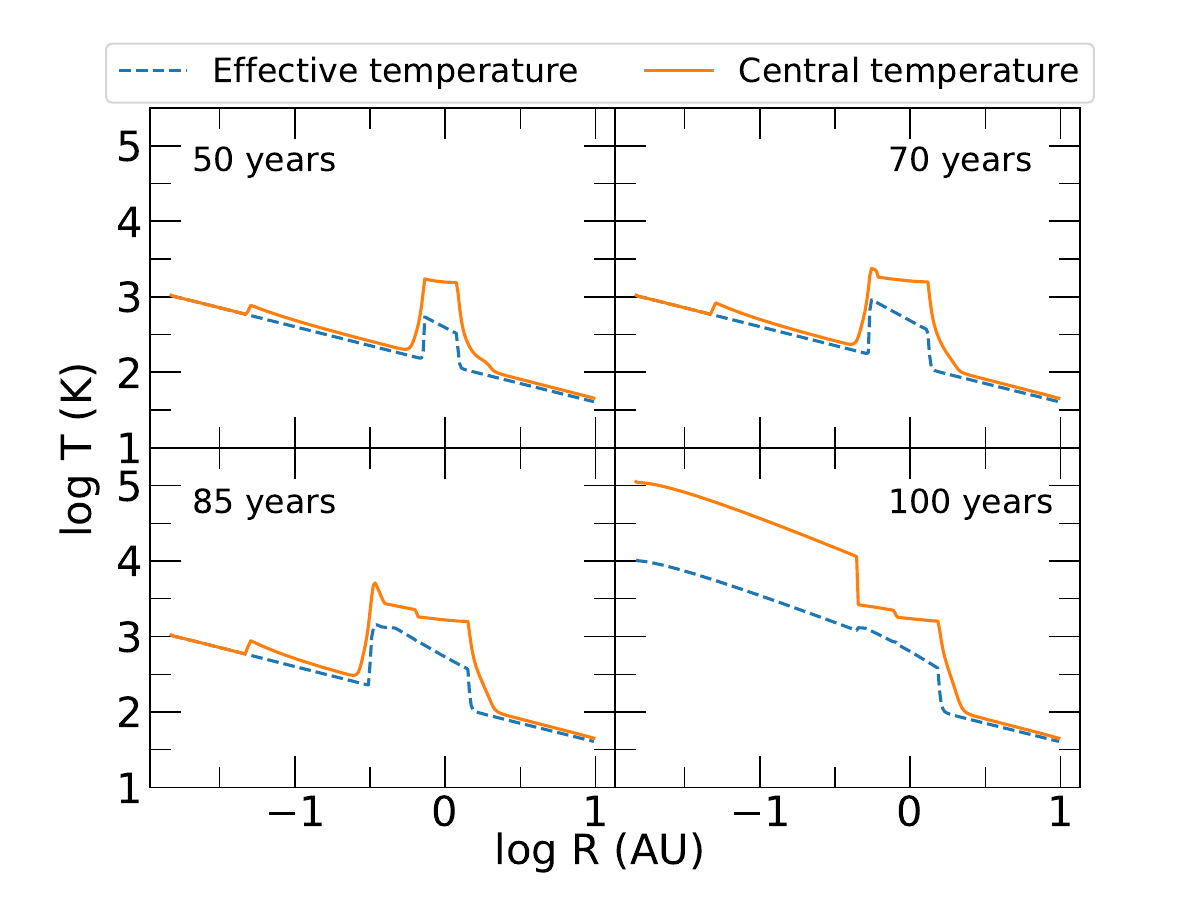}
    \caption{Evolution of the "large" (FU Ori-like) outburst model, showing the surface (effective) temperatures (blue
    dashed curves) and the central temperatures (orange solid curves), respectively, at four times after the initial triggering. }
    \label{fig:jacobtempplot}
\end{figure}

We computed several radiative transfer
models for the time evolution of two different
outbursts.
For models 2A - 4B we adopt
the inner disc temperature distributions of
the ``FU Ori-like'' model of \cite{cleaver2023} which culminate in
a large outburst with peak accretion rates
$\sim 10^{-4} \msunyr$. To achieve
such a large outburst, 
it was
triggered with a temperature perturbation
at the relatively large radius
of 1 au; consequently, the accretion front takes many years
to reach the innermost disk and
result in a burst at optical wavelengths.
We selected snapshots
at $t = 50,\, 70,\, 85,$ and $100$~yr 
after outburst triggering  for radiative
transfer calculations
(see Figure \ref{fig:jacobtempplot}).
This enables us to follow
the evolution of the SED from precursor phases
to a "final" phase in which
the disc SED is essentially that of a standard steady-state,
optically-thick disc. 

\begin{figure}\centering
    \includegraphics[width=\columnwidth]{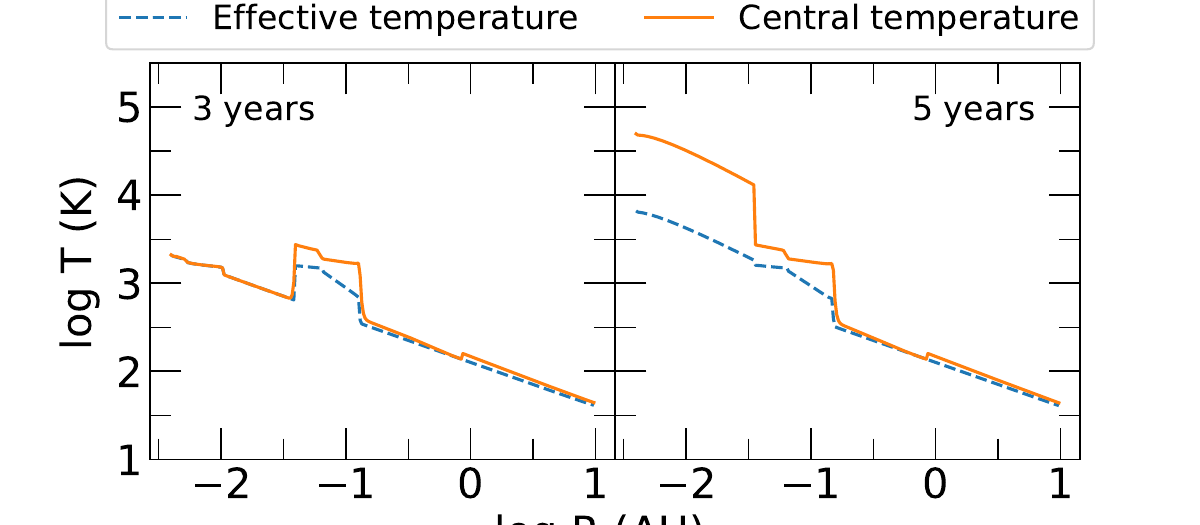}
    \caption{Evolution of the "small" (Gaia 17bpi-like) outburst model, showing the surface (effective) temperatures (blue dashed curve) and the central temperatures (orange solid curve), respectively, at two times after the initial triggering. }
    \label{fig:jacobtempplotGAIA}
\end{figure}

The second set of model calculations take the disc temperature distributions
from
the simulations of \cite{cleaver2023} 
modeling the smaller outburst of Gaia17bpi. This outburst was triggered at a smaller radius of
0.1 au, to achieve lower accretion rates and faster evolution.
Two times are chosen; one just before
the optical outburst at $t = 3$ yr, and another at $t = 5$~yr when the accretion
rate was $4 \times 10^{-7} \msunyr$
(Figure \ref{fig:jacobtempplotGAIA}).

We calculate envelope models
for two mass infall rates:
a `high' infall rate of $1 \times 10^{-5} M_\odot / yr$, based upon the maximum of `typical' rates of \citet{zhu2009}, and a lower rate of $1 \times 10^{-6} \msunyr$.
(The important quantity for the
radiative transfer solutions is the
density structure, not the infall rate which depends upon the free-fall velocity. Our models thus correspond
to true infall rates
$\mdot (true) = \mdot (model)
(M_* / \msun)^{-1/2}$.)
The envelope extends to a maximum radius of $6000$ au, and 
we adopt an envelope outflow cavity within the streamline corresponding to $\theta_0 = 35^{\circ}$.
These properties are typical of models for
protostellar envelopes
\citep[see, e.g.][]{furlan2016}. 
In all but one case we assume a centrifugal radius $r_c = 30$~au. This is based on observations finding
median protostellar disc radii $\sim 30-50$~au
\citep{tobin20}. 

For the central star we adopt a stellar
spectrum 
with effective temperature of 3500 K 
and $\log g = 3$ from the 
BT-NextGen\footnote{\url{http://svo2.cab.inta-csic.es/theory/newov2/index.php?models=bt-nextgen-agss2009}} models
\citep{allard11,allard12}. 
We further adopt two sets of stellar parameters, one for each outburst type: a typical
pre-main sequence stellar radius
$R_* = 2 \rsun$ and a stellar mass $M_* = 0.44 \msun$ in the case of the ``FU Ori'' model, and a smaller radius of
$R_* = 0.8 \rsun$ and stellar mass $M_* = 0.2 \msun$ for the Gaia 17bpi model. The mass only enters in
setting the conversion from densities
needed for the continuum radiative transfer to mass infall rates by changing
the free-fall velocity (see above).

The disc parameters were the same for
all cases:
$R_0 = 10$~au, $R_T = 30$~au, and $\Sigma_0 = 750\, {\rm g\, cm^{_2}}$,
with the fiducial disc scale height set to
$H_0 = 0.809$ au at $R_0$.
The resulting disc mass is large
($0.16 \msun$), consistent with the possibility of GI triggering of large outbursts \citep[][and references therein.]{cleaver2023}.

Figure \ref{fig:jacobtempplot} shows four snapshots of the disc temperature structure from the FU Ori-like simulation. 
The viscous heating is triggered at
$\sim 1$~au, and then the burst propagates inward as material 
accretes. 
Before the outburst reaches the inner disc,
the central temperatures tend to thermostat at roughly 1500 K because we assume that dust
is sublimated above that value (see discussion in \citealt{cleaver2023}) and \S \ref{sec:discussion}). 
When the outbursting 
region reaches the inner disc 
radii, dust is sublimated and the
gas enters the gas opacity-dominated,
thermally-unstable regime, causing temperatures to rise dramatically.
In this stage the temperature distribution is essentially that of the standard
steady, optically thick accretion
disc, except that there is no 
turndown in T(R) at small radii 
because the simulations do not
employ the zero-torque inner
boundary condition \citep{shakura1973}.

Figure \ref{fig:jacobtempplotGAIA}
shows the temperature structure
of the Gaia17bpi model disc at
two times. In this case the outburst
was triggered at a much smaller
radius, 0.1~au, to make a better
match of the calculated lag time
between optical and infrared 
outburst signatures observed by
\cite{hillenbrand2018}.
The overall evolution is similar
to that of the FU Ori-like models,
except that it is much faster.

\subsection{SEDs for FU Ori-like outburst}

 \begin{figure}
    \includegraphics[width=\columnwidth]{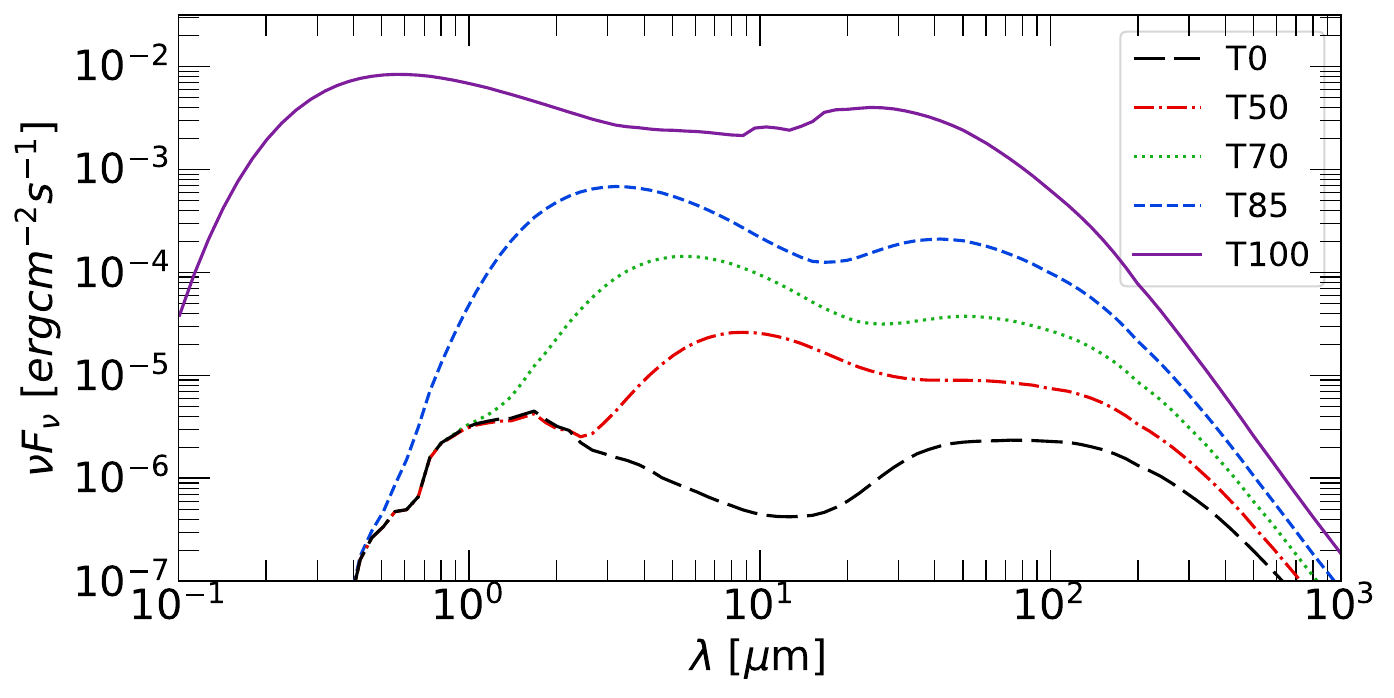}
    \caption{A comparison between the spectral energy distributions at an inclination of $0^{\circ}$ for Model 1, with no outburst, Models and 2A - 2D of the FUor outburst
    sequence. The SED becomes brighter with increasing time.}
    \label{fig:timecomp}
\end{figure}

Fig.~\ref{fig:timecomp} shows the evolution of the SED for the FU Ori-like
outburst, viewed at $i = 0^\circ$
(down the cavity) to
illustrate how the sequence of disc temperature distributions shown in Figure \ref{fig:jacobtempplot} affect the SEDs without including envelope
extinction.
As the outside-in outburst progresses, the disc emission gets brighter and the peak of the SED gradually shifts to shorter wavelengths until the
accretion front reaches the inner
disc, at which point the disc becomes
very much hotter and luminous.
The mid- to far-infrared emission
similarly becomes brighter and bluer
as the outburst evolves \citep[e.g.,][]{furlan2016,fischer24}.

\begin{figure}
    \includegraphics[width=\columnwidth]{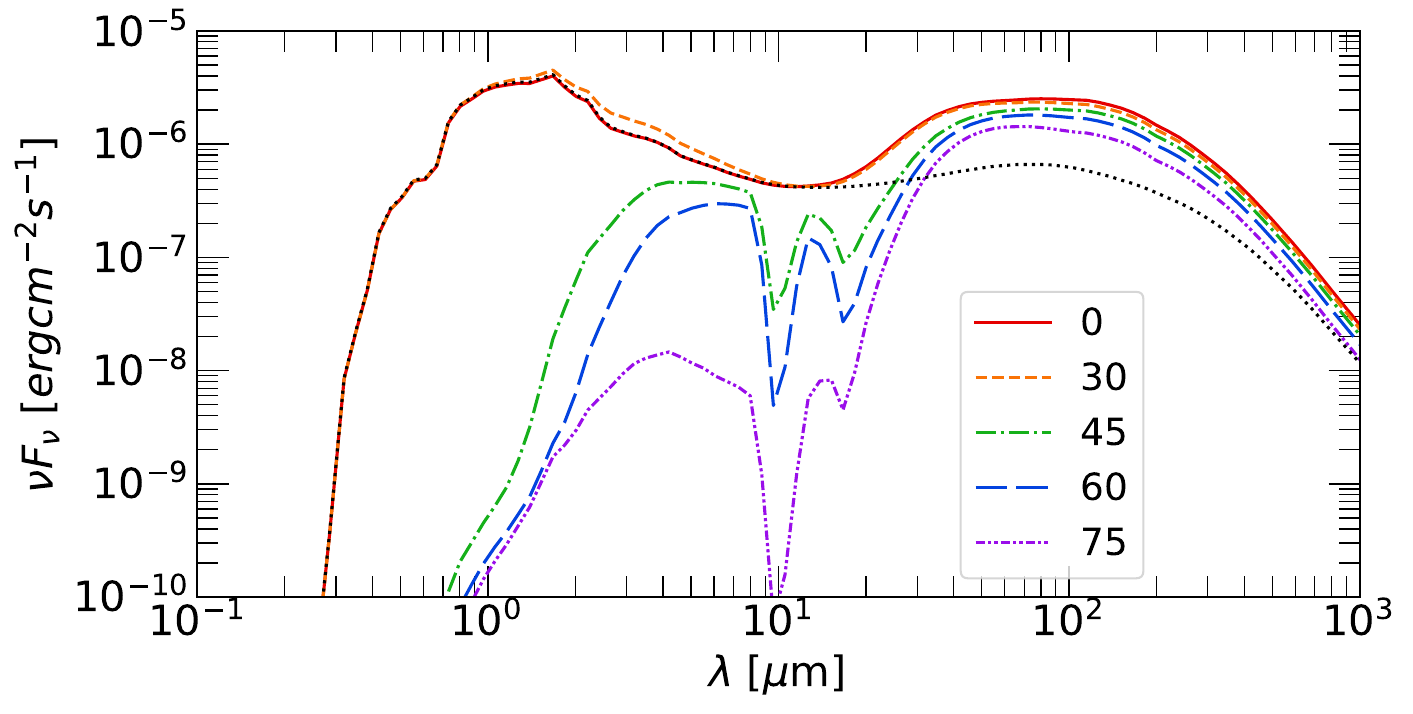}
    \caption{SEDs for Model 1 (star + disc + envelope), with an infall rate of $10^{-5} \msunyr$, with no outburst.
    The SEDs are fainter at higher inclination.
    The black dotted curve is 
    the star and disc without envelope viewed at $i = 0^{\circ}$.}
    \label{fig:staranddiskonly1e5ENVNOHEAT}
\end{figure}

\begin{figure}
    \includegraphics[width=\columnwidth]{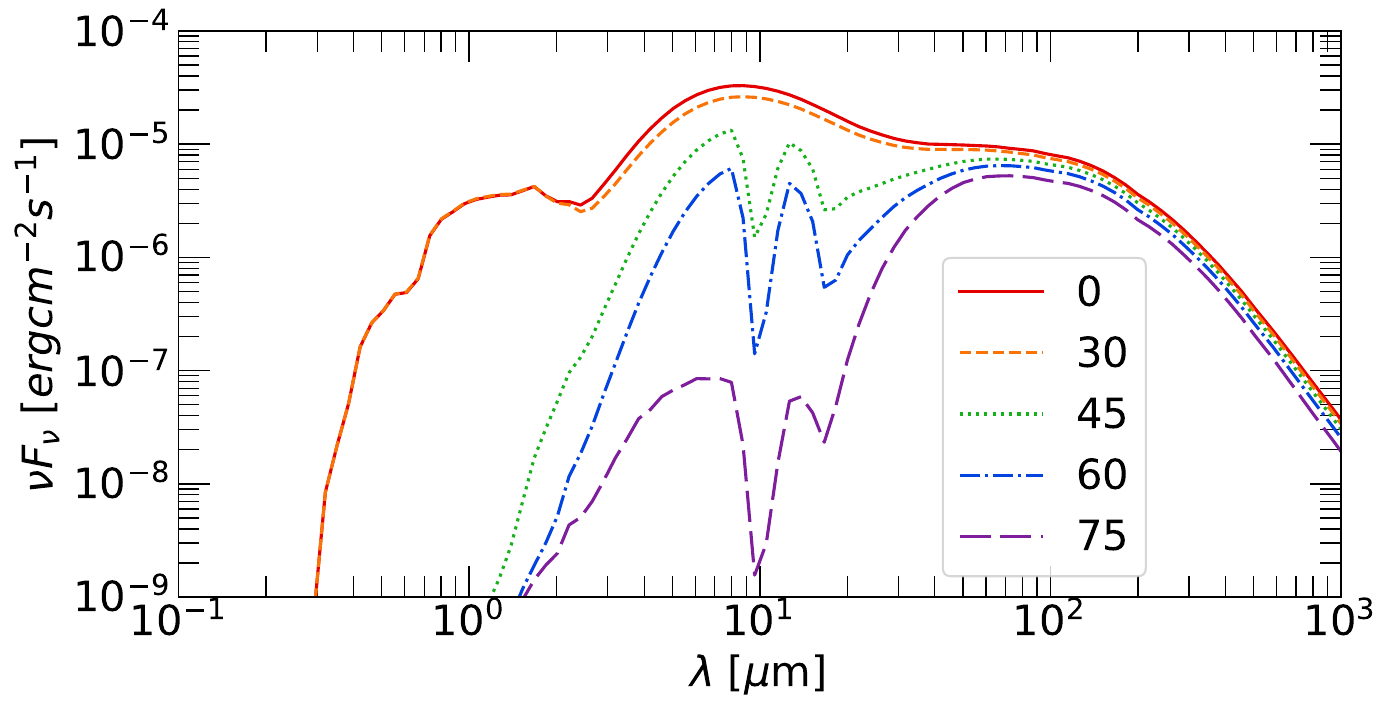}
    \caption{SEDs for Model 2A, with infall rate of $10^{-5} \msunyr$, $r_c = 30$~au, at a time $t = 50$ yr after
    triggering. The infrared precursor produces a disc SED that peaks at around $8-10 \mu$m.} 
    \label{fig:T50fulltreat}
\end{figure}

\begin{figure}
    \includegraphics[width=\columnwidth]{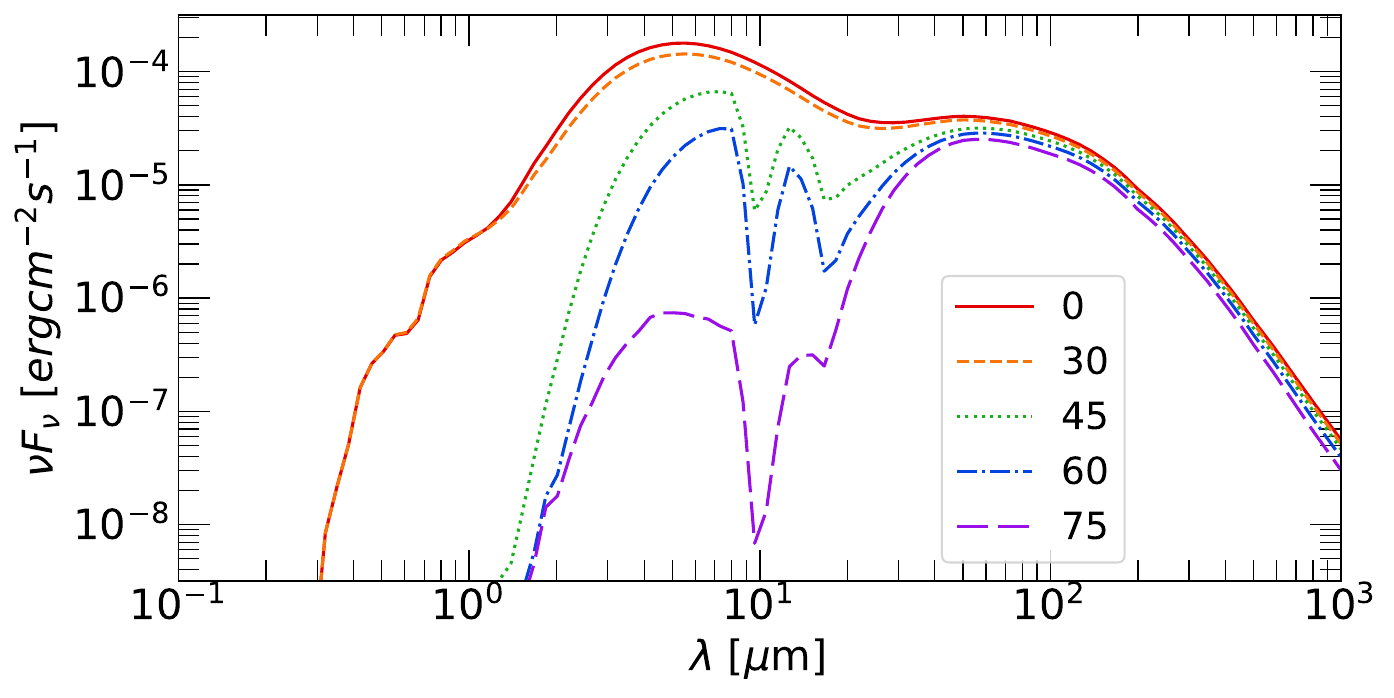}
    \caption{SEDs for Model 2B, with the
    outburst at $t = 70$ yr. Other parameters are the same as
    as Fig.~\ref{fig:T50fulltreat}.
    The precursor SED becomes brighter and
    bluer as the burst propagates inward.}
    \label{fig:T701e5finalplot}
\end{figure}

\begin{figure}
    \includegraphics[width=\columnwidth]{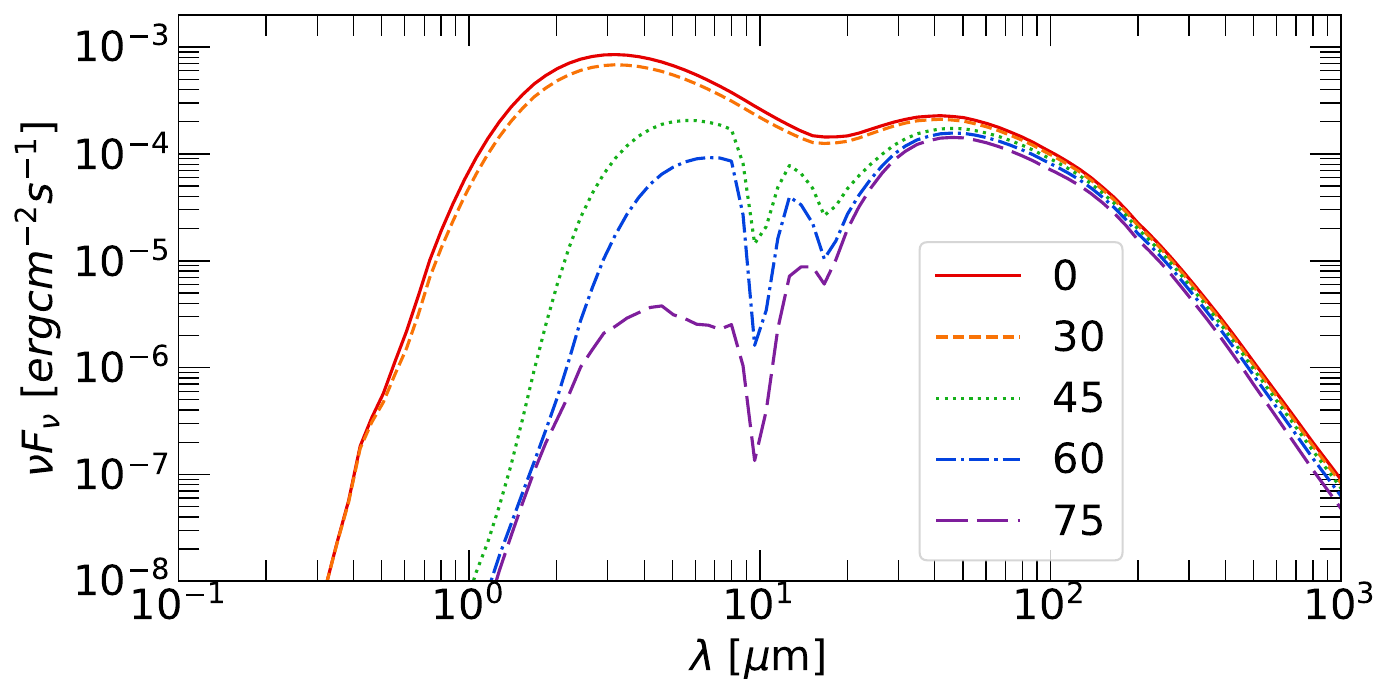}
    \caption{SEDs for Model 2C, with the
    outburst at $t= 85$ yr.
    Other parameters are the same as in 
    Fig.~\ref{fig:T50fulltreat}. }
    \label{fig:T85fulltreat}
\end{figure}

\begin{figure}
    \includegraphics[width=\columnwidth]{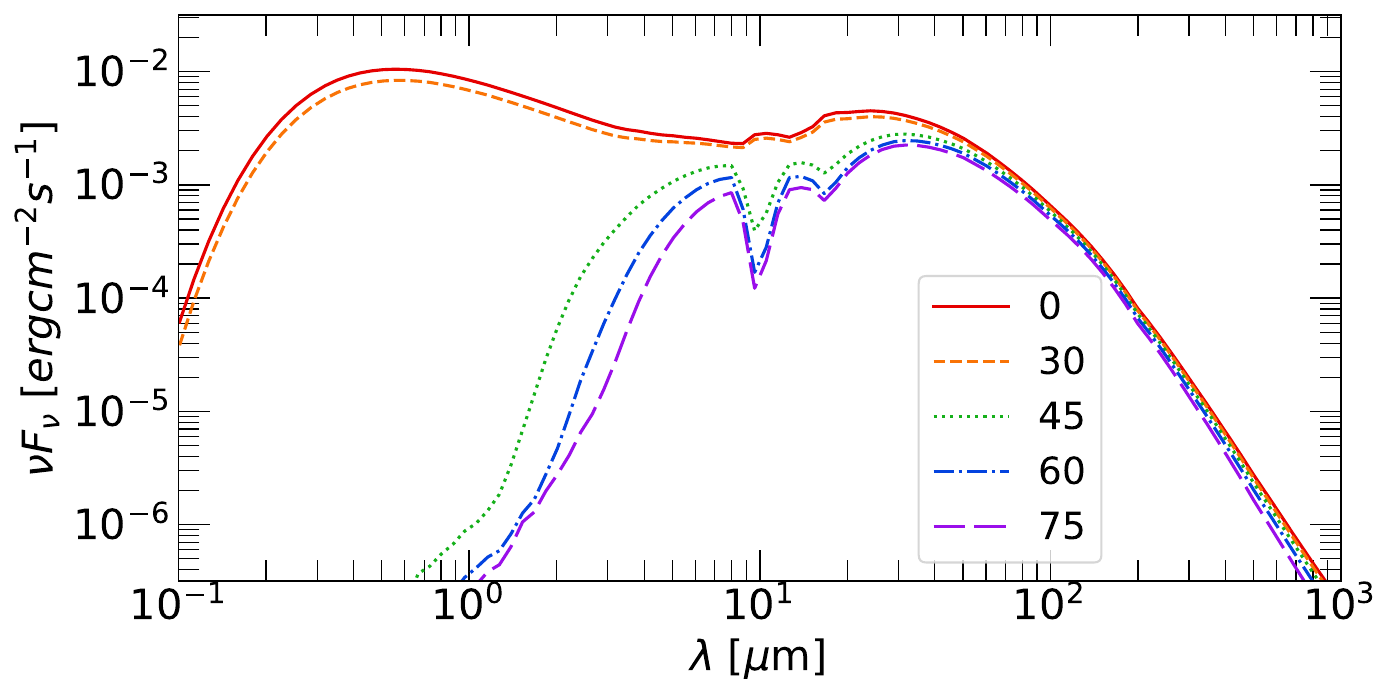}
    \caption{SEDs for Model 2D, with the
    outburst at $t = 100$ years. 
    In this case the outburst has reached the inner disc, with much higher peak temperatures and luminosities. At this 
    stage the disc SED is essentially that of a standard steady accretion model.} 
    \label{fig:T1001e5finalplot}
\end{figure}

\begin{figure}
    \includegraphics[width=\columnwidth]{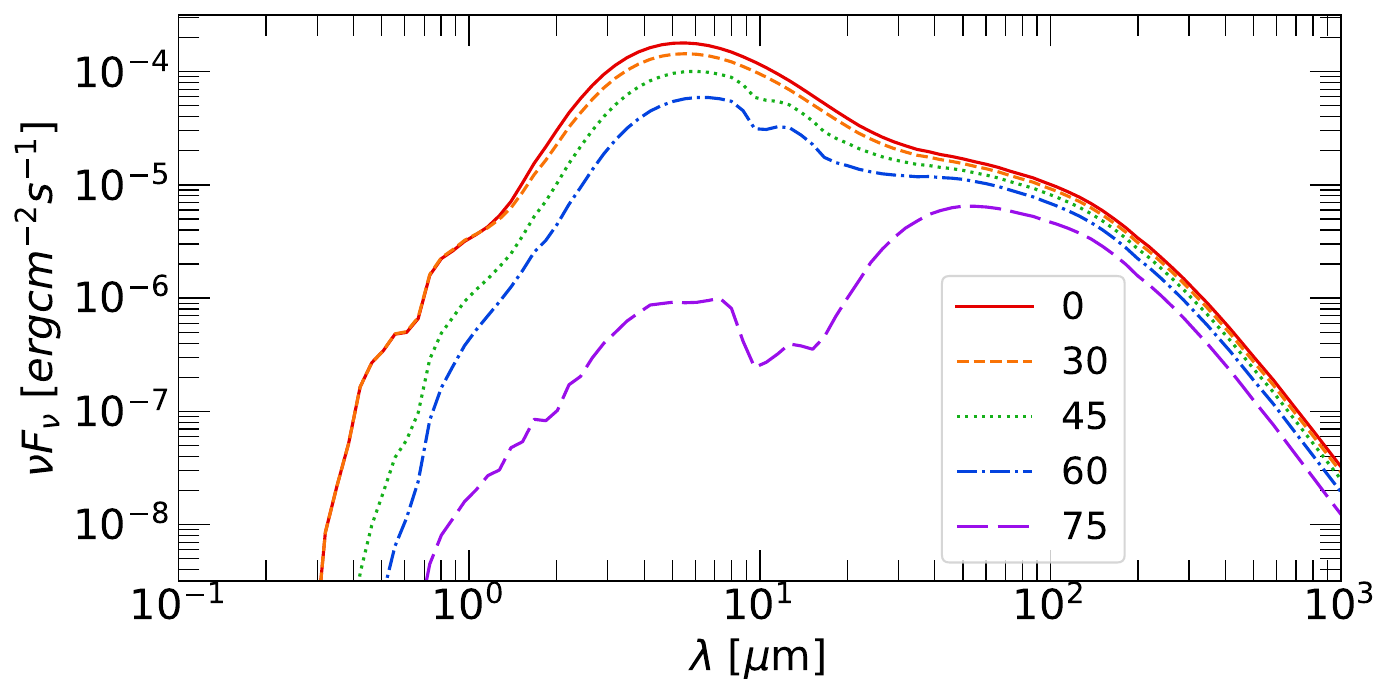}
    \caption{Spectral energy distributions for Model 3 (star + disc + envelope), 
    at $t = 70$~yr with an infall rate of $10^{-6} \msunyr$ (Table 1).}
    \label{fig:T701e6finalplot}
\end{figure}





\begin{figure}
    \includegraphics[width=\columnwidth]{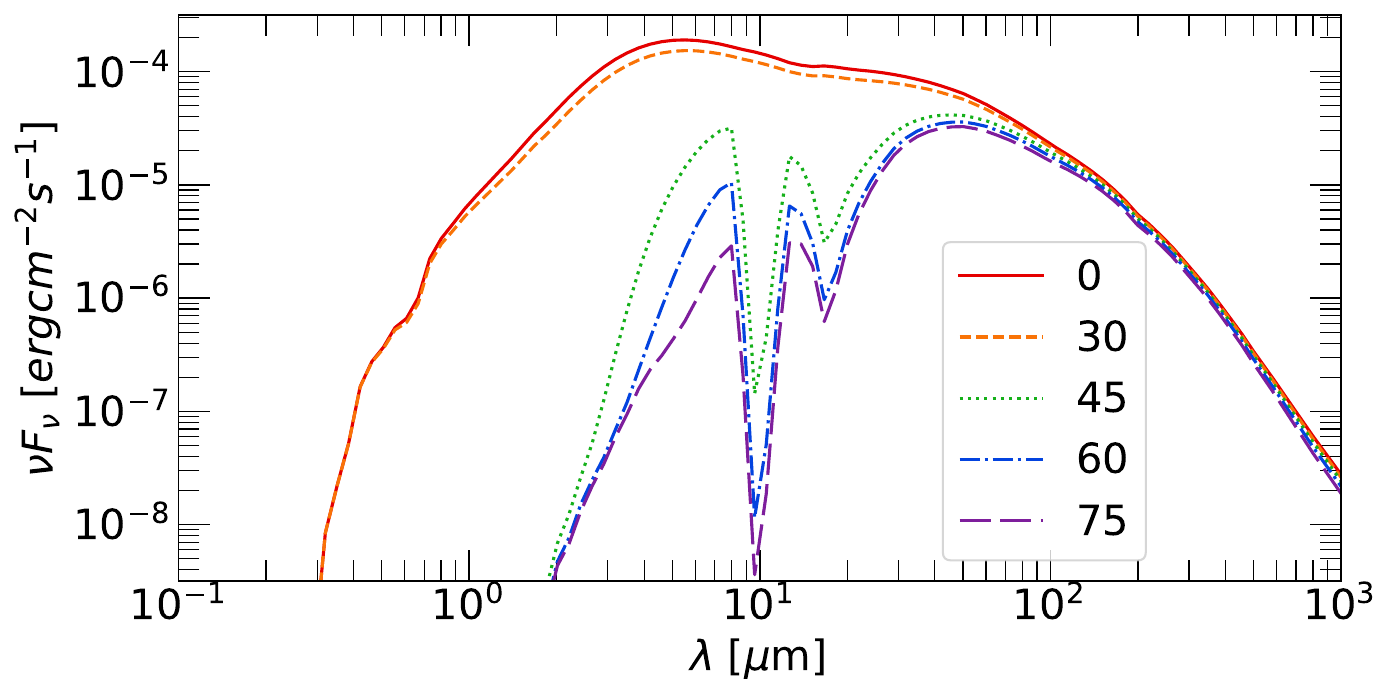}
    \caption{Spectral energy distributions for Model 4. This case is equivalent to Model 2B (Fig.~\ref{fig:T701e5finalplot}), but with a much smaller centrifugal radius $r_c = 3$ ~au (see text). }
    \label{fig:T70rc3AU}
\end{figure}



\begin{figure}
    \includegraphics[width=\columnwidth]{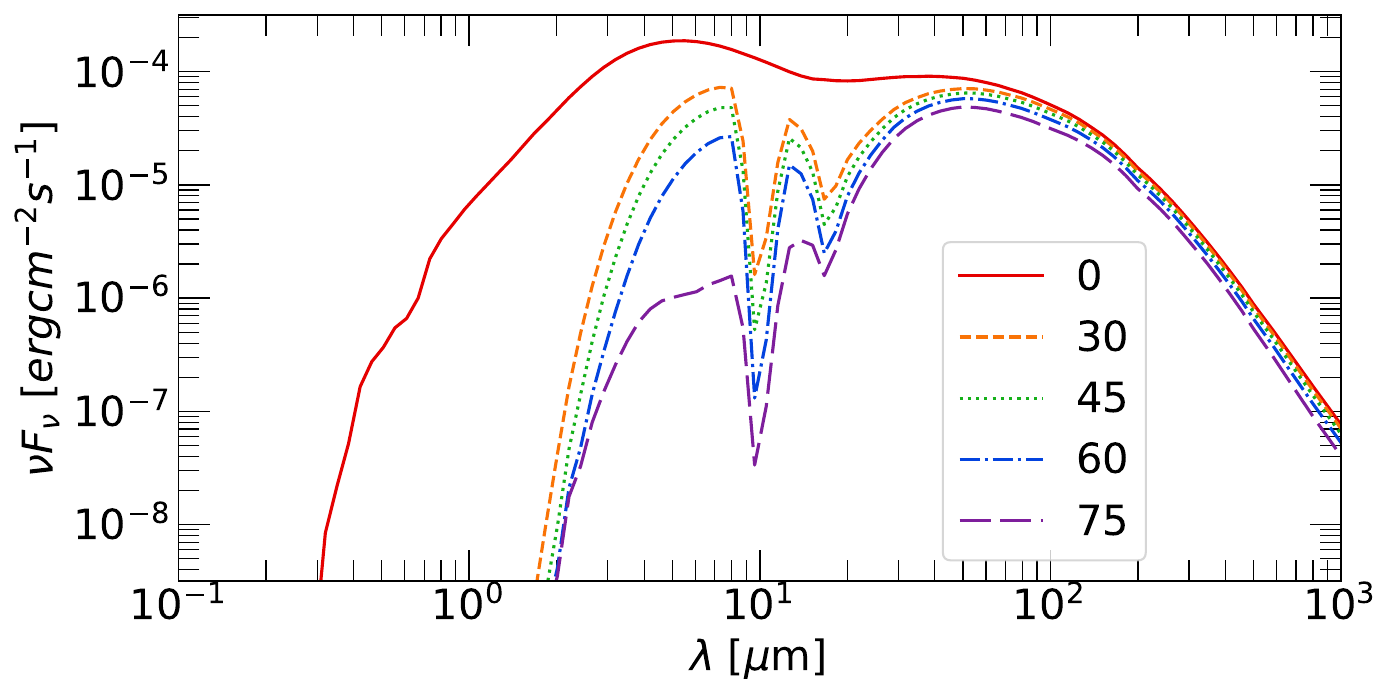}
    \caption{Spectral energy distributions for  Model 2B', with the same parameters as Model 2B but with the cavity opening angle reduced from $35^{\circ}$ to $10^{\circ}$
    (see text). }
    \label{fig:T70env1e5CAV10}
\end{figure}

Fig.~\ref{fig:staranddiskonly1e5ENVNOHEAT} shows SEDs of Model 1 with only the star, envelope, and non-outbursting disc
at different inclinations.
This is intended to illustrate
what the system would look like
pre-outburst.
At low inclinations, viewing the system through the cavity, the optical to mid-infrared spectrum of the
system is clearly evident.
The large far-infrared excess due to the envelope is only weakly dependent upon inclination.
Because we adopt a relatively
geometrically-thick disc, the absorption features are due to 
a combination of disc and envelope extinction at $i = 75^{\circ}$.
For reference,
the black dotted curve shows the star plus disc without envelope viewed at $i = 0^{\circ}$.

Figures~\ref{fig:T50fulltreat} through
Fig.~\ref{fig:T1001e5finalplot},  models 2A - 2D, show the time evolution of the FUor-type outburst sequence for the envelope parameters of $\mdot = 10^{-5} \msunyr$ and $r_c = 30$~au at different inclinations.
In the following we use the
term ``infrared precursor'' 
or simply ``precursor'' 
to refer to the increase in
infrared kmission before
the outburst is seen at optical wavelengths.

While the precursor is clearly
evident when the system
is viewed down the outflow
cavity, its signature in the SED is much
less evident if at all for
higher inclinations.
At $i = 45^{\circ}$
and $i = 60^{\circ}$
the emission at $\lambda \sim 5-10\, \mu$m (outside of the silicate absorption features)
is somewhat brighter
than the far-infrared flux
from $\sim 30-100 \, \mu$m,
in contrast to typical
protostellar SEDs, in which the
$\sim 5-10 \, \mu$m flux is fainter than
the far-infrared emission
(compare with \ref{fig:staranddiskonly1e5ENVNOHEAT}; also \citealt{furlan2016} and \citealt{fischer24}).
This inverted flux ratio might be
a sign of the precursor,
although it could be difficult to
distinguish from an infrared companion (see futher discussion in \S \ref{sec:discDetect}).
However, at $i = 75^{\circ}$ the extinction due to the
envelope and disc is too large
to reveal any precursor emission.

Because the precursor emission
peaks at $\sim 3-10 \mu$m, it is not
surprising that the amount of
envelope extinction is crucial
to identifying the early outburst
in the SED.
Figure \ref{fig:T701e6finalplot}
shows the outburst at $t = 70$~yr,
for the same parameters as
model 2B (Figure \ref{fig:T701e5finalplot}, but
for the lower infall rate of
$10^{-6} \msunyr$). The precursor
is clearly detectable at all but the highest inclinations due to
the lower envelope extinction.

Conversely, envelope extinctions
that are higher than in cases
2A-2D will make it very difficult
to detect precursors unless viewed
down the outflow cavity.
The optical depth toward the central
source in these envelope models
scales approximately as
$\tau \sim \mdot r_c^{-1/2}$ (see \S \ref{sec:discDetect}),
so increasing this combined
parameter will render any
precursor undetectable.
As an example, Figure \ref{fig:T70rc3AU} shows Model 4, with the same parameters as Model 2B except for a smaller centrifugal radius of 3 au. The increased extinction
results in only a faint hint of the precursor at $i = 45^{\circ}$ and none at higher inclinations.

The extinction toward the central
source in these models also depends somewhat on our choice of
outflow opening angle.
This is because we define our cavities 
by the asymptotic angle $\theta_{out}$ of a streamline, which means that the infalling
dusty mass lands on the disc
at radii
\begin{equation}
r_{in} \geq \sin^2 \theta_{out} \, r_c \,.
\label{eq:rin}
\end{equation}
Figure \ref{fig:T70env1e5CAV10}
shows Model 2B', with the same parameters as Model 2B 
but changing 
$\theta_{out}$ from $35^{\circ}$
to $10^{\circ}$. 
The extra extinction makes the precursor
essentially undetectable, which is not surprising given that the inner edge of the infalling material
at the disc has moved inward from
7.5~au to 0.9 au (equation \ref{eq:rin}). Note that this
means some infalling material
is landing on part of the outbursting region of the disc.
Note also that for both
the small $r_c$ case
(Figure \ref{fig:T70rc3AU}) as well as this small opening angle case the near-infrared excess becomes undetectable unless viewed
down the cavity.

This discussion only addresses the detection of the
infrared precursor in the SED. However, the evolution of the
precursor can be traced indirectly through the overall increase
in the system luminosity as the outburst propagates inward,
even at high line-of-sight extinctions. We address the time variability
as a function of the wavelength of observation in \S 4.

\subsection{SEDs for Gaia 17bpi outburst}

The other outburst case we consider uses the models of 
\cite{cleaver2023} for
Gaia 17bpi, which exhibited a much smaller outburst than the
FU Ori case, $\sim 1-6 \times 10^{-7} \msunyr$
\citep{hillenbrand2018,rodriguez2022}. Such smaller outbursts are more common than the large FUor bursts \citep[e.g.,][]{fischer23,contreras24}, so we use these models to evaluate the detectability of these less powerful, shorter timescale events.

\begin{figure}
\centering
\includegraphics[width=\columnwidth]{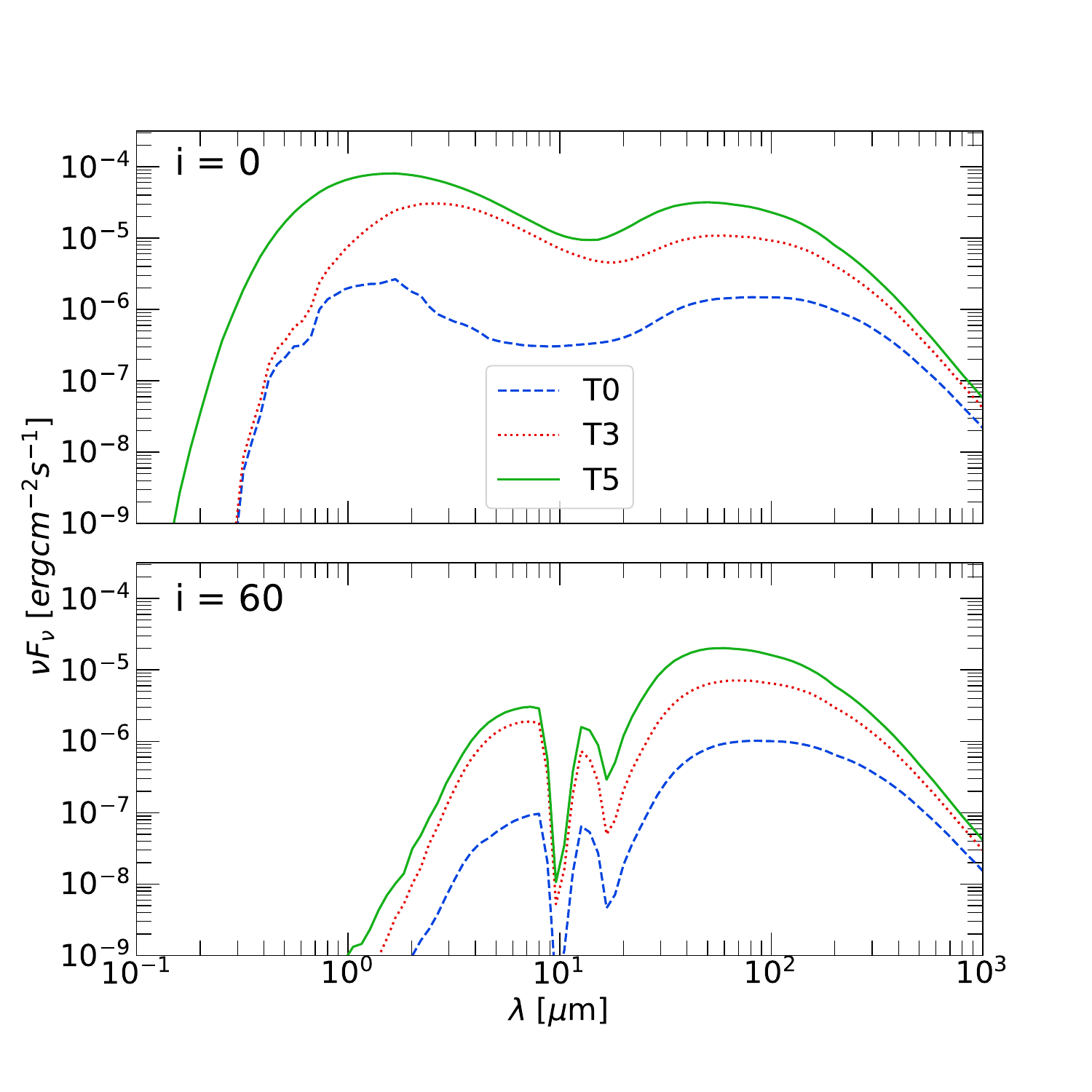}
    \caption{Spectral energy distributions for Models 5A - 5C of the Gaia outburst. Three times are compared, at 0, 3, and 5 years, corresponding to a non-outbursting, precursor, and peak outburst. The upper panel compares the three at a 0-degree inclination, while the lower compares the three at a 60-degree inclination. } 
    \label{fig:T3burstcomp}
\end{figure}

The models for Gaia 17bpi by
\cite{cleaver2023} assumed
that the outburst was triggered
at radius of $0.1$~au, compared with the 1~au triggering of 
the FU Ori-like outburst
(Figure \ref{fig:jacobtempplot}).
This difference results
in much faster evolution
and a much shorter time
lag between the infrared
precursor and the optical
outburst.
Because the accretion front
does not have far to propagate
inwards before maximum light,
the precursor is only seen
as a peak in the SED at 
$\sim 2-3 \, \mu$m about
two years before the
SED becomes that of a steady
accretion disc, as shown
in in the $i=0$ sequence
in Figure \ref{fig:T3burstcomp}.

\begin{figure}
    \includegraphics[width=\columnwidth]{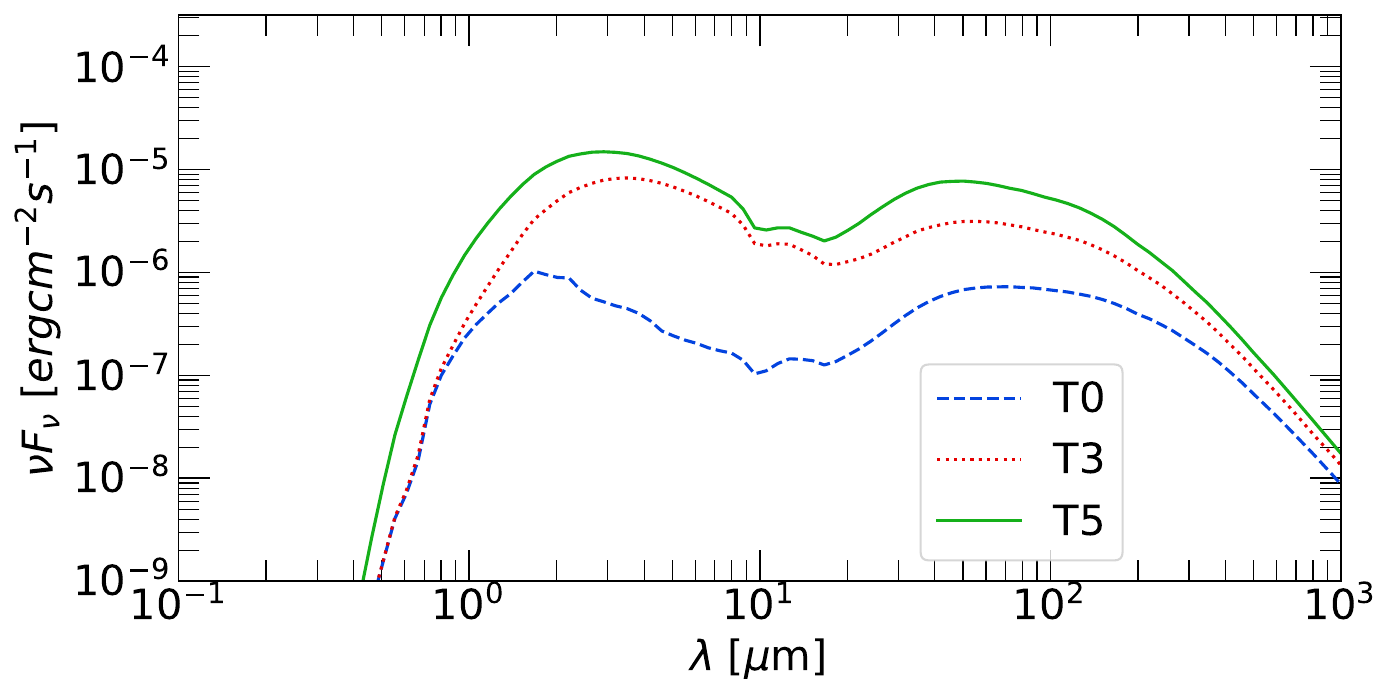}
    \caption{SEDs for the Gaia disc outburst case but with
    $\mdot = 10^{-6} \msunyr$ (Models 5D,E,F), observed at an inclination of $i = 60^{\circ}$.} 
    \label{fig:GAIAIN601e6}
\end{figure}

The lower panel of Figure
\ref{fig:T3burstcomp} shows that
the precursor is not detectable
for $i = 60^{\circ}$ 
at $10^{-5} \msunyr$ and
$r_c = 30$~au. Unlike Model 2B,
the precursor emission is considerably
redder, where the extinction is
larger. For the lower infall rate
of $10^{-6} \msunyr$, the precursor is
detectable at $i=60^{\circ}$ (Figure \ref{fig:GAIAIN601e6}),
but it would be difficult to
differentiate the source spectrum
from a steady disc with a lower
maximum temperature.

\section{Discussion}
\label{sec:discussion}

\subsection{Detectability and evolution of outburst}
\label{sec:discDetect}

The outburst simulations predict that
during the precursor phase, before the optical rise, the disc SEDs exhibit a peak in the $2-10\, \mu$m region,
indicating that the temperature of the hot region of the disc is 
around $\approx 1000$~K.
There are two reasons for this.
One is the choice of our trigger
central temperature of $\sim 1300$~K, with a somewhat
lower surface or effective
temperature. The trigger temperature
is particularly important for
low, inner disk outbursts
such as those of Gaia17bpi, because
the burst does not propagate
very far inward before optical
maximum light. 

For large, long-lasting
outbursts such as the FU Ori-type case, another effect
is more important for
the evolution of the precursor: the
thermostatic effect caused by the
assumed sublimation of dust at
$T \sim 1500$~K. The central temperature in the hot region is maintained by viscous dissipation 
with radiative trapping that results in a higher central temperature than at the surface. 
If the temperature rises above
1500~K, the disappearance of dust results in a large decrease in the 
opacity - which then reduces the radiative trapping and cools the interior. 
The net effect of this thermostatic
behavior is that the hot region's effective temperature remains very roughly at $\sim 1000$~K during the precursor
phase. 

As the large burst propagates inward,
the effective temperature of the hot region does not remain precisely constant but becomes somewhat hotter (Figure \ref{fig:jacobtempplot}). This causes the SED peak to move from
$\sim 10\, \mu$m at early phases
to $\sim 3\, \mu$m 
as the
overall outburst brightens (see Fig.~\ref{fig:timecomp}).
The SED only changes dramatically
when the burst reaches small radii.
At this point
the viscous heating becomes too large for dust to survive, and the dependence of the gas opacity on temperature 
triggers a thermal instability resulting in
much higher temperatures
(Fig.~\ref{fig:jacobtempplot}, bottom right panel).
In this phase the SED has essentially
the shape of the 
standard steady, optically-thick accretion disc
spectrum (Figure \ref{fig:T1001e5finalplot}).

The envelope modifies the
disc SEDs by both absorbing and reeemitting radiation from the central
regions. Given that the peak temperature in the disc is $\sim 1000$~K for many years
in the large outburst case,
the precursor emission peak occurs at a wavelength where the envelope mostly absorbs. 
In these models the
optical depth along a particular line of sight to the central regions scales as
\begin{equation}
    \tau_{\lambda} \propto k_{\lambda} \mdot r_c^{-1/2}\,,
\end{equation}
where $k_{\lambda}$ is the opacity.   As illustrated in
our ``base'' case, 
with $\dot{M} = 10^{-5} \msunyr$ and $r_c = 30$ au,
a hint of the disc SED peak
is present at $i = 45^{\circ}$
but not apparent at larger
inclinations. For $\dot{M} = 10^{-6} \msunyr$ and the same centrifugal radius,
the reduction in envelope
optical depth of a factor
$\sim 3$ results in the
disc SED peak being very detectable, except at the highest inclinations, where the disc absorption becomes dominant. Similarly,
for $\dot{M} = 10^{-5} \msunyr$ and $r_c = 3$ au,
the increase in optical depth 
prevents detection of the precursor SED peak.

The other effect of the envelope is to reradiate
the central source emission
at long wavelengths. For the larger values of centrifugal radius, and our chosen cavity
properties, the far-infrared peak of the disc SED is well displaced in wavelength from
the precursor peak
(Models 2A-D).
However, for model B', with $r_c = 3$ au, the infalling
dust extends much closer to the star and becomes much hotter, such that the envelope emission begins to merge in wavelength with that of the precursor as seen down the cavity. 
In this case the presence of
the precursor SED peak may be difficult to identify with
confidence.

Another possible problem in identifying an infrared precursor would be misidentification due to presence of an embedded binary companion that is only
detected at infrared wavelengths.
An example of this is the Z CMa system, which consists of an FU Ori object well-detected at optical wavelengths, with an infrared companion separated by
only $0.1 "$ that dominates the system luminosity \citep{koresko91}.
The SED of the joint system
is qualitatively similar to that of the fiducial model.
Ruling out an infrared companion would require interferometry or possibly
detection of the companion in
scattered light \citep{whitney93}.


Obviously it is much easier
to detect the outburst precursors predicted by
these models when viewing along
an outflow cavity. We have chosen a relatively large
opening angle 
which may overestimate
the probability of finding
objects with favorable inclinations. Another potential problem is
the presence of external
extinction. In modelling 
the SEDs of Orion protostars,
\cite{furlan2016} 
inferred large external
extinctions for many sources.
Absorption of $A_V \gtrsim 10$ would obscure any
potential precursor signal
otherwise unaffected by
envelope extinction.

Despite the dificulties of clearly 
detecting a precursor photometrically
in highly extincted systems, observing
the
time evolution of long-wavelength
emission could provide independent evidence
of precursor outburst evolution.
In an analysis of the  
Spitzer/IRAC
Candidate YSO (SPICY) Catalog for variables in the inner galaxy
\citep{kuhn21}, supplemented where possible with data from the VVV project \citep{smith25}, \cite{contreras25} attempted to
identify candidate FUors from slow,
``linear'' increases or decreases in
light rather than the usual search
for quick, large-amplitude bursts.
These types of light curves could
correspond to the slow decay phase of outburst, as seen in FU Ori \citep{herbig77}, or a slow rise in accretion. The rising light curves 
could represent an inside-out 
burst \citep{bell94} as observed
for V1515 Cyg \citep{herbig77}.
Alternatively, linear rising light
curves might also be the signature
of outside-in precursors.
With spectroscopic followup of a small subsample, \cite{contreras25} were able to confirm that genuine
FUors are identifiable from these
light curve classifications, potentially greatly expanding the census of outbursting young objects,
and further observations of systems slowly rising in brightness could yield
insights into burst triggering even
without detailed SED signatures.

\subsection{Comparison with prior work}

\begin{figure}
    \includegraphics[width=\columnwidth]{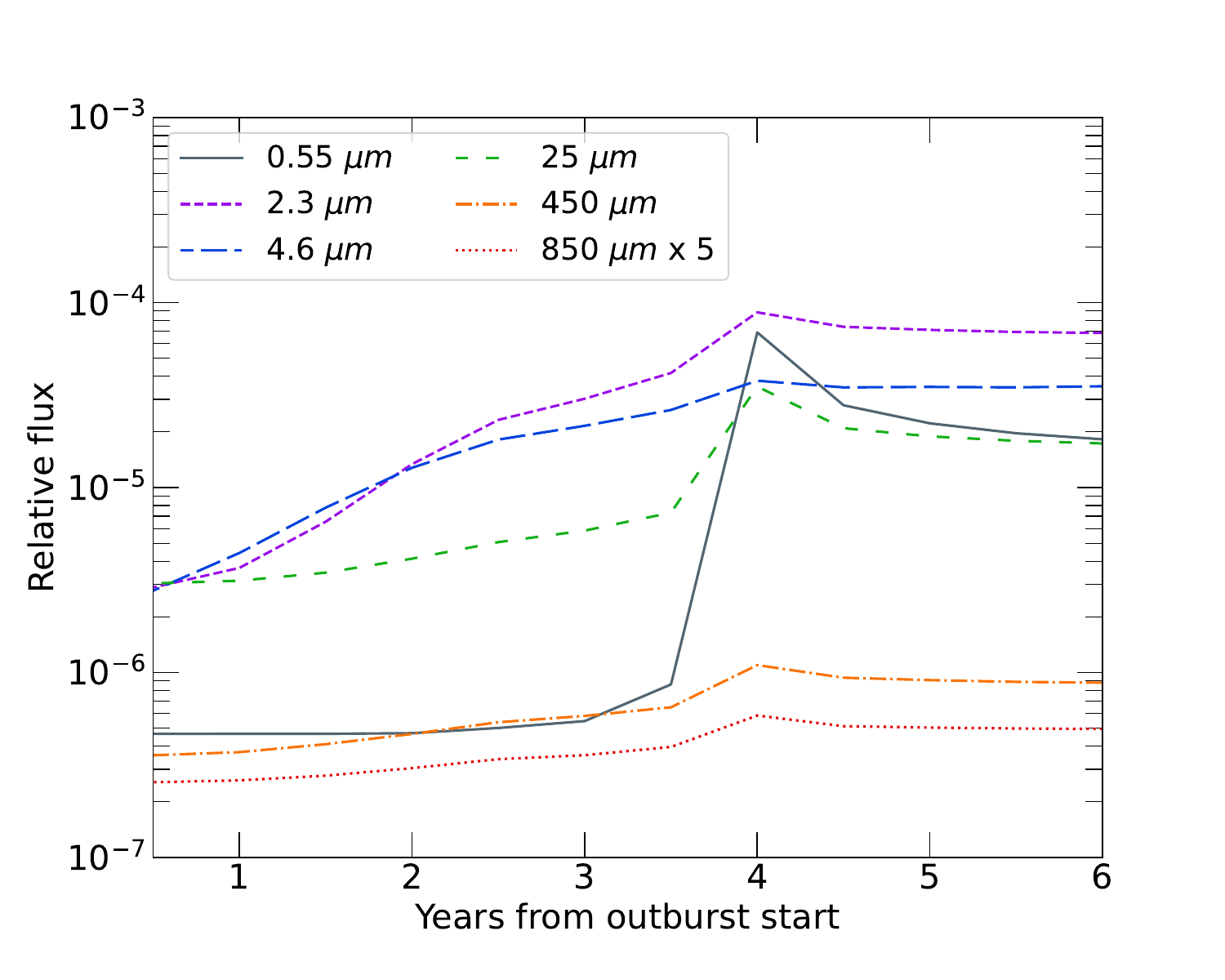}
    \caption{Light curves of relative flux for the Gaia-type outburst, demonstrating its time evolution for a variety of wavelengths. Note that the parameters are identical to Models 5A - 5C, observed at an inclination of $0^{\circ}$. 
    } 
    \label{fig:timedepGAIA}
\end{figure}


\begin{figure}
    \includegraphics[width=\columnwidth]{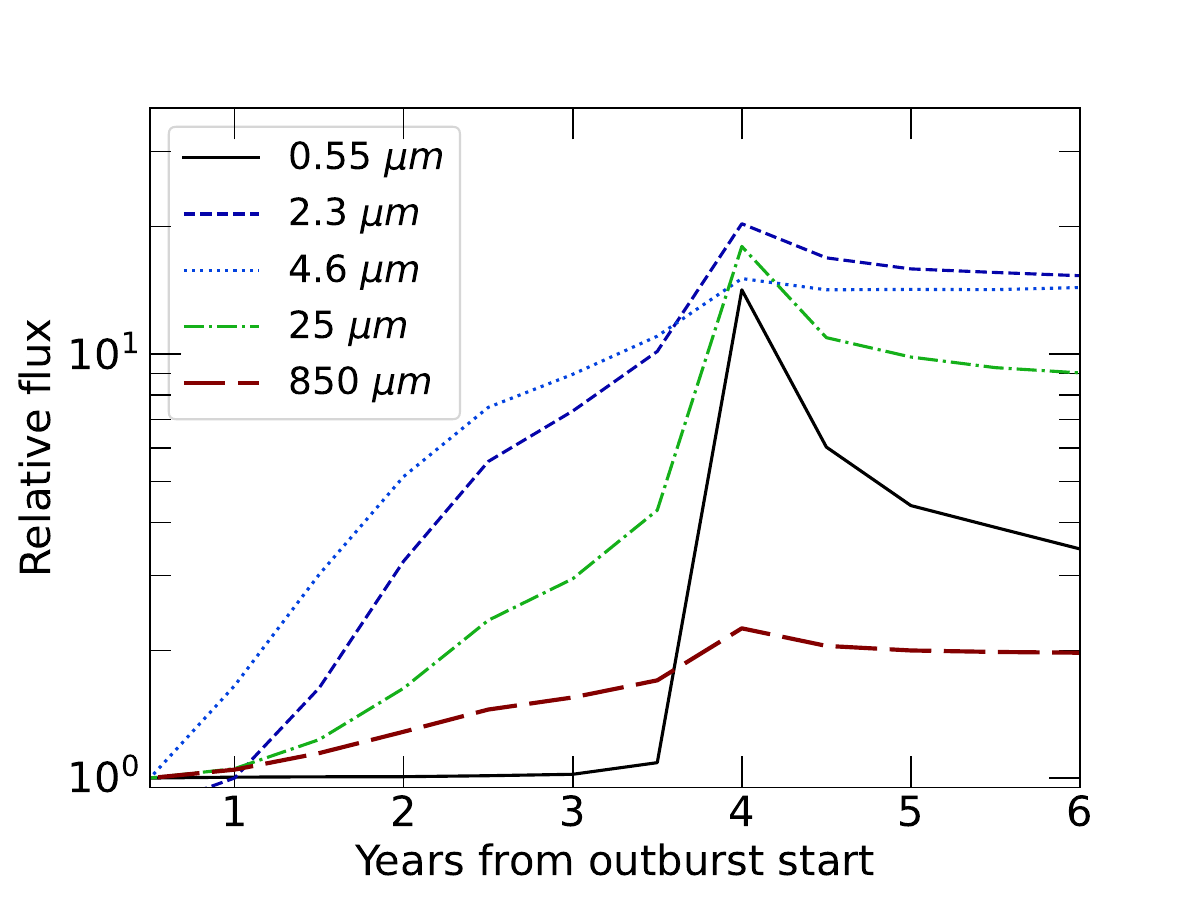}
    \caption{Flux evolution at five wavelengths for
    the Gaia17bpi outburst
    (Models 5D,E,F), observed through the envelope at a typical inclination of $60^{\circ}$, but now normalized to the flux in the pre-outburst state.} 
    \label{fig:relativeflux3wavIN60}
\end{figure}

\begin{figure}
    \includegraphics[width=\columnwidth]{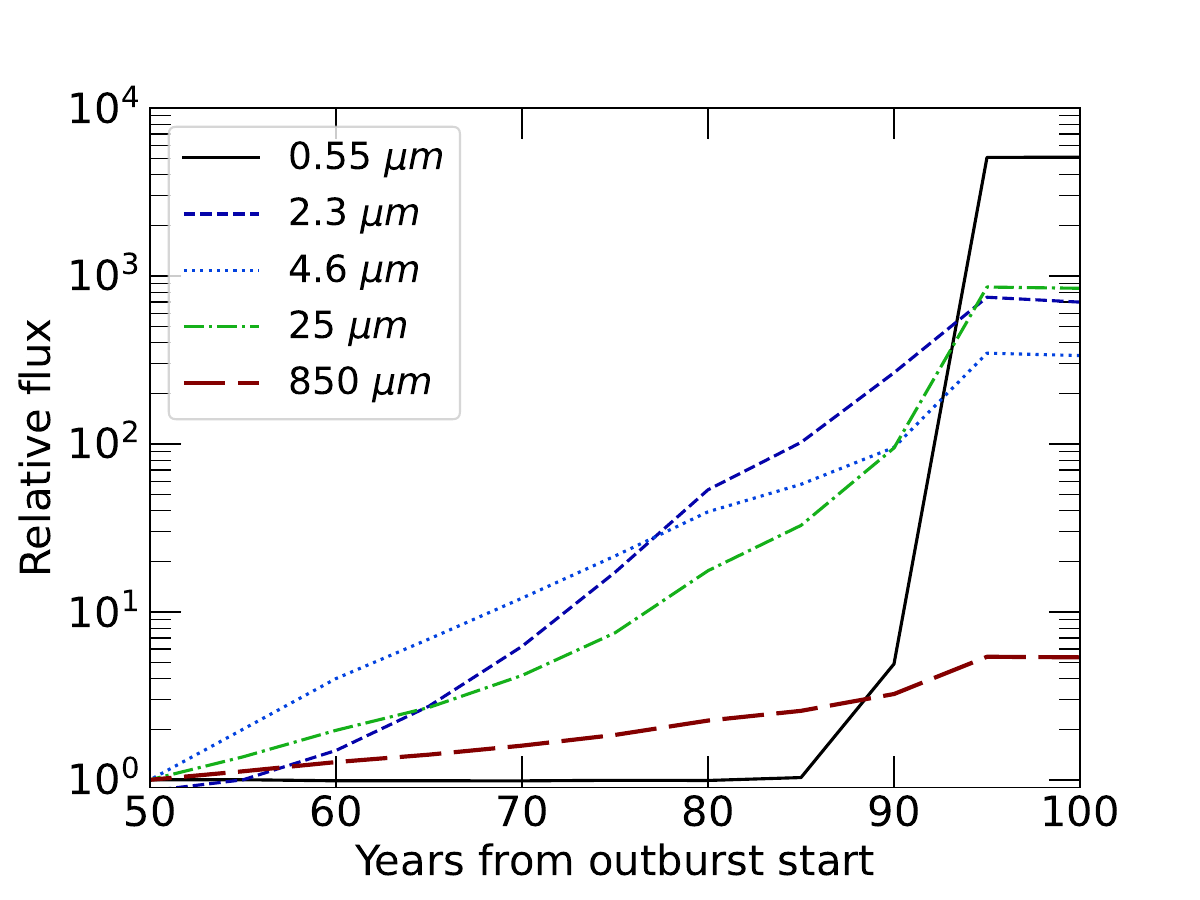}
    \caption{Evolution of the fluxes at the same five wavelengths as in Figure \ref{fig:relativeflux3wavIN60}, but for the FU Ori-like outburst sequence (Models 2A,B,C,D), normalized to the 
    flux at 50 years after initial triggering, and observed through the envelope at $60^{\circ}$.}
    \label{fig:relativeflux3wavIN60FUORI}
\end{figure}

The previous study most relevant
for this work is that of 
\cite{fischer24}, which explored the most favorable wavelength ranges to
detect outbursts of highly extincted
protostars.  Their envelope and disc
models were taken from the results of \cite{furlan2016} for 86 Class 0 
protostars in Orion. 
Fischer et al. computed SEDs for these models with increases in luminosity ranging from 10 to 100 times the
original, and then analyzed the fractional changes at various wavelengths relative to the true change in luminosity. 
Consistent with earlier, more limited studies
\citep{baek20,macfarlane19,contreras20}, Fischer et al. concluded that far-infrared measurements most
closely represented the intrinsic
luminosity changes, and that far-IR and longer wavelength observations 
were preferred over studies at shorter wavelengths strongly affected by extinction.

Our aim in this paper is different from Fischer et al. in that we consider detectability 
and time evolution of
the infrared precursor predicted by
models of outside-in bursts.
As the signature of the model
is a peak in disc emission in the
$2 - 10 \,\mu$m wavelength range,
our focus is on detection at shorter wavelengths than those favored more
generally by Fischer et al.
In addition, we focus on the time
evolution of the outburst. As shown in Figures \ref{fig:timedepGAIA}
and \ref{fig:relativeflux3wavIN60}, the
precursor in the Gaia models is strongest at $ 4.6\, \mu$m, whereas the 
$25 \, \mu$m signal is somewhat smaller, and very weak in the submm region, as found by \cite{fischer24}.

Qualitatively similar flux evolution is shown in 
Figure \ref{fig:relativeflux3wavIN60FUORI} for the FU Ori-like outburst; the main difference is that the
increase in the fluxes is slower and ultimately larger. \cite{contreras25}
found potential FUors with
``linear'', slow rises in brightness at $3.4~\mu$m and $4.6 ~ \mu$m (WISE W1 and W2 bands), increasing by $\sim 1-2$ magnitudes
over $\sim 10$~yr.
roughly consistent with
the flux evolution shown in Figure \ref{fig:relativeflux3wavIN60FUORI}. The
rise timescale in the simulations is set by
the location of the initial trigger and the
adoption of a viscosity
parameter $\alpha = 0.1$.
While the choice of $\alpha$ was set
by matching burst evolution to observations of Gaia17bpi, it must
be emphasized that full
global magnetohydrodynamic
simulations, which show much more complex behavior than can
be captured with a one-dimensional, alpha
viscosity treatment
\citep[e.g.,][]{zhu20,roberts25},
are needed to make more
robust predictions.

Once the outburst reaches peak
luminosity, most of the radiation
originates near the central protostar.
Thus at this stage the simulations
are only moderately different
from the Fischer et al. treatment,
which assumes that all the radiative energy originates from the central protostar with
a fixed effective temperature, varying the protostellar radius to
achieve differing system luminosities.
Viewed down the outflow cavity, the
SED of the disc will look different from the star, but at other inclinations and with high enough envelope/external extinction, the source spectrum will not matter as the
system SED will be dominated by
emission from the dusty envelope.
Another difference is that while we use the same density structure 
as that of \cite{furlan2016}
(equation \ref{eq:envdens}), we use 
a different cavity shape, which generally
makes little difference to SEDs
(although it is important for modeling scattered light images).
Finally,
a disc has a different angular distribution of radiation than that of a spherical star (see Appendix), but this is likely of secondary importance considering other uncertainties in envelope and cavity structures.

Our results show that, for outside-in accretion bursts,
near-infrared monitoring can be useful in detecting 
precursors except in the most heavily-extincted systems.
In any case our simulations
reinforce the conclusion of
\cite{fischer24} that far-infrared
and longer wavelengths can reliably
detect outbursts in protostars.
Because the youngest systems
will be the most embedded, long-wavelength monitoring can probe
the onset of outbursts
in the earliest
stages of protostar formation.


\section{Conclusions}
\label{sec:conc}

We have computed SEDs for
protostars to explore
the effects of infalling envelopes 
on the detectability of 
of outside-in accretion outbursts.
Specifically, we consider
models of outside-in propagation
of accretion bursts that predict mid-infrared precursors that
are in principle detectable
before the optical light burst.
We find that for envelope infall
rates $\gtrsim 10^{-5} \msunyr$ to disk radii
$\sim$ 30 au, the 
peak of the precursor emission is difficult to discern due to high extinction
unless the system is viewed down an outflow cavity.
However, at more modest infall rates
$\lesssim 10^{-6} \msunyr$, 
the mid-infrared peak is detectable, except at high inclinations where disc occultation/extinction is important.
We find agreement with the results of \cite{fischer24} that
mid-infrared studies are more sensitive to outbursts than submm and mm observations. However, in addition our 
outburst models predict that
the $\sim 2-10\, \mu$m range is
more sensitive to the outburst
precursor. Monitoring at long
wavelengths presents
intriguing possibilities of
of detecting precursor luminosity increases, particularly in the infrared \citep[e.g.,][]{contreras25}, but even submm/mm
monitoring can provide useful
constraints.
Detection of such precursors, and measurements of
the lag time between infrared and optical bursts,
can be important in constraining
the radial distance at which
the outburst starts \citep{cleaver2023}.
Our results may help inform observing strategies for improving 
the characterization 
of protostellar accretion.

\section*{Acknowledgements}

We are extremely grateful to Jacob Cleaver and Fiona Han for allowing access to their work, as well as aiding us in its operation and understanding. We would also like to thank Cornelis Dullemond for answering our questions over correspondence, and Nuria Calvet for providing detailed opacity tables.
LH was supported in part by
NASA Emerging Worlds grant 80NSSC24K1285.
Software: RADMC3D \citep{dullemond2012}, SPECTRES \citep{carnall2017}, MATPLOTLIB \citep{hunter2007}.

\section*{Data Availability}

The models and associated data will be provided upon reasonable request to the authors. 



\bibliographystyle{mnras}
\bibliography{bibliography} 



\appendix

\section{Inclination-dependent apparent luminosity}
\label{app:inclination}

\begin{figure}
    \includegraphics[width=\columnwidth]{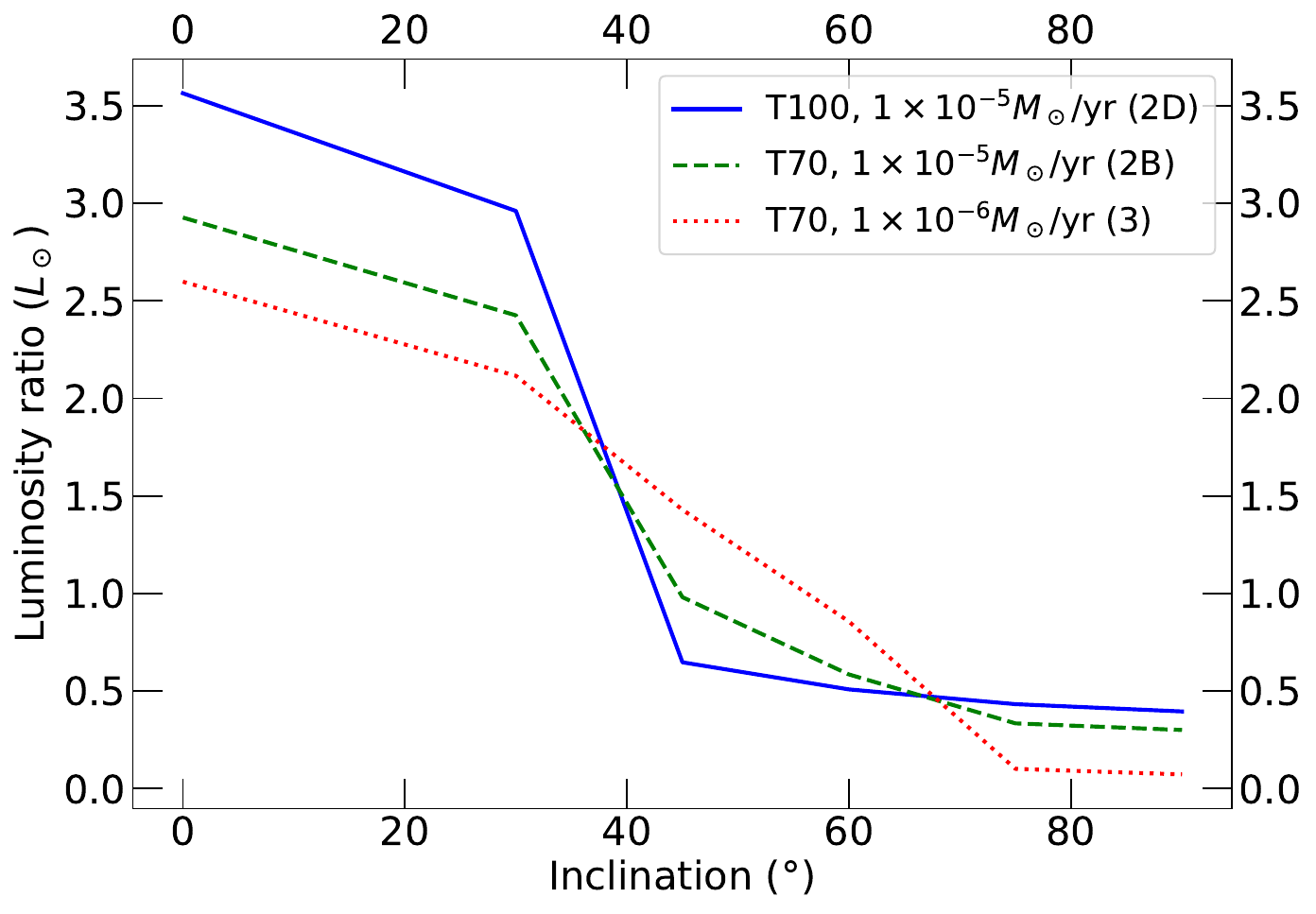}
    \caption{The inclination dependence of luminosity for Models 2B, 2D, and 3, including full stellar, disc, and envelope components. The flashlight effect is clearly apparent, as is its dependence on envelope infall rate. }
    \label{fig:inclination}
\end{figure}

The radiation from discs is 
anisotropic; in addition, protostellar
envelopes generally are not spherical,
which can cause important variations in spectral
appearance as a function of
inclination.
\cite{whitney03} explored the effects of non-spherical geometry on the SEDs,
images, and apparent luminosities of
protostars.
Their envelope models were essentially the same as ours but
with somewhat different outflow cavity parameters.
Whitney et al. found that pole-on viewing angles might result in
overestimating true source luminosities
by a factor $\sim 2$ due to the
``flashlight effect'' \citep{yorke1999}, with the
envelope geometry in effect channelling radiation out lower extinction regions. Conversely,
Whitney et al. found that
more edge-on systems might appear underluminous by a factor of 1.5 
(or more if occulted by the disc; see their Figure 10).

While \cite{whitney03} did include
disc emission, most of the accretion luminosity arose from inside the
dust destruction radius; and this radiation was assumed to come from
the central star. As a result, the
radiation from central regions in
the Whitney et al. models is
more spherically symmetric than
in our calculations. As an example of the effect of the anisotropy of
disc emission, in Figure \ref{fig:inclination} we show 
what the inferred luminosity
would be relative to the true value
for Models 2B, 2D, and 3).
As expected, observations at
low inclinations yield inferred
luminosities brighter than the true
value by factors $\sim 3$, higher
than
that found by Whitney et al. due to
the anisotropy of disc emission.
Conversely, as found by Whitney et
al., objects viewed at a median
inclinations, $\sim 60^{\circ}$,
will appear underluminous, by
a factor of two or more.


\bsp	
\label{lastpage}
\end{document}